*This is the author's version of the work. It is self-arhived at Arxiv. The definite version was published in: Nguyen Duc A., Dahle Y., Steinert M., Abrahamsson P. (2017) Towards Understanding Startup Product Development as Effectual Entrepreneurial Behaviors. In: Ojala A., Holmström Olsson H., Werder K. (eds) Software Business. ICSOB 2017. Lecture Notes in Business Information Processing, vol 304. Springer, Cham, https://doi.org/10.1007/978-3-319-69191-6_15*

# The application of an entrepreneurial theory to study software development in startups

Anh Nguven Duc [1,2], Yngve Dahle[1], Martin Steinert[1], Pekka Abrahamsson[1,2]

[1] Norwegian University of Science and Technology, NO-7491 Trondheim, Norway
[2] Software Startups Research Network http://softwarestartups.org

**Abstract.** Software startups face with multiple technical and business challenges, which could make the startup journey longer, or even become a failure. Little is known about entrepreneurial decision making as a direct force to startup development outcome. In this study, we attempted to apply a behavior theory of entrepreneurial firm to understand the root-cause of some software startup's challenges. Six common challenges related to prototyping and product development in twenty software startups were identified. We found the behavior theory as a useful theoretical lens to explain the technical challenges. Software startups search for local optimal solutions, emphasize on short-run feedback rather than long-run strategies, which results in vague prototype planning, paradox of demonstration and evolving throw-away prototypes. The finding implies that effectual entrepreneurial processes might require a more suitable product development approach than the current state-of-practice.

**Keywords:** effectuation, entrepreneurial behavior theory, software development, software startups, prototyping, empirical study

## 1 Introduction

The software industry has witnessed a growing trend, where software products are developed by startup companies with limited resources and little operating history. With the advancement of technology development, it seems that everyone with a business idea, a website and a pitch can launch a new company. However, not so many business ideas are realized as concrete prototypes. Furthermore, even a smaller portion of prototypes is transformed into commercialized products. It is difficult to repeat successes, as startups operate in chaotic situations, where the links between startups' behaviors and their effects are often not detectable [1].

Decisions made by entrepreneurs is the direct force leading to the success or failure of the startup [3]. Startup's unique characteristics, i.e. dynamic, bootstrapping and multiple-influenced environments, make the decision-making tasks for entrepreneurs are different for project managers in more established companies [1]. Entrepreneurs often have to make decisions with little information about market, customer and product, and whether they will be accepted [2]. Entrepreneurial literature offers several ways to understand the startup's decisions and behaviours [3, 4, 8]. One approach is the behaviour theory of entrepreneurial firms, which assumes the effectuation approach when developing startups' business [4]. Recent ideologists [5, 6, 7] encourage the co-development of business and product in startups. The combination of the two line of thoughts inspires us to explore the effectual behavior of startups from product development aspect. We are interested in understanding how the theory of entrepreneurial behaviors could help to explain the challenges faced during the product development. Our research question is "*How are theories of entrepreneurial behaviors applicable to explain for startup product development process?*"

The paper is organized as follows; firstly, we present related work about a behavior theory of entrepreneur firm (Section 2). Then, we describe our research methodology (Section 3). After that, findings are presented (Section 4). Finally, we will discuss and conclude the paper (Section 5 and 6).

## 2 Behavioral Theory of the Entrepreneurial Firm

Entrepreneurship literature is intensive on understanding the formation, development and influencing factors to startups. There has been an increased attention on the effectuation theory in explaining entrepreneurial behaviours [8]. Effectuation processes take a set of means as given and focus on selecting between possible outcomes that can be realized [8]. Alternatively, entrepreneurial firms are seen as heterogeneous, bounded rational entities [4]. In the face of environmental uncertainty, therefore, these bounded rational firms form expectations based on available means and information. Dew et al. proposed a behavioral theory of the entrepreneurial firm (BTEF) [4]. Assuming entrepreneurs as an effectual unit, Dew et al. [4] propose four constructs related to entrepreneurial decision-making:

- Means-driven transformation: startup companies tend to be effectual, and available resources drive their action. Effectual action involves transforming extant means into new possibilities, including new problems of interest. Transformation processes are actor-centric, as who comes on board determines goals, not vice versa. The transformation is appeared as a search activity, aiming at solving pressing problems rather than developing long-run strategies.
- Docility: conflict and difference among stakeholders is avoided through stakeholder docility, goals are residual of the process. Simon et al. defined docility as "*the tendency to depend on suggestions, recommendation, persuasion and information obtained through social channels, as a major basic of choice*" [9]. The decisions made by startups, for instance, can be done by in cooperating other's ideas and not necessary by going through conflict resolution.

- Leveraging contingency: avoiding uncertainty by short run feedbacks, but also encouraging surprise. For startups, even '*bad*' surprises can be leveraged to provide new means and new opportunities. Actions emphasize commitment and contingency, not choice and determinacy.
- Technology of foolishness: insulation from learning sought through allowing experimental actions with regard to affordable lost. The technology of foolishness allows startups to relax the primacy of functional rationality, to temporarily suspend intentionality, and promote the openness to new actions, objectives and understandings.

## 3 Research Approach

We conducted this study by using a multiple-case study design with software startup as a unit of analysis [10]. Contacts for startups were searched via four channels, (1) startups within professional networks of papers' authors, (2) startups in the same town with the authors, (3) startups listed in the Startup Norway website and (4) the Crunchbase database. Twenty startups were eventually selected for investigation. The startup cases represent different startup phases, from prototyping to commercialization and scaling. Application domains range from marketplace, education, ecommerce, transportation, and Internet-of-Thing. Regards to software development approaches, startups with five or more people mostly adopt Agile and iterative software development. The sample is dominated by Norwegian software startups, with small teams and bootstrap financing models. We do not consider other types of startups, for example, internal cooperate startups, venture capital invested startups, and USA-based startups.

The major data collection instrument is semi-structured interviews. The interviewees were focused on exploring startup's decision making and their behaviors related to their business and product development. The interview guideline is published online[1]. We used a thematic analysis to analyze the data, a common technique for identifying, analyzing, and reporting conceptual themes found from qualitative data [17]. To support the data analysis, we used a tool namely NVivo 11[2], to code, and to categorize such codes in higher order levels, representing different technical challenges when going from ideas to commercialized product. Several theoretical frameworks were considered, such as Cynefin model [11], boundary spanning object theory [12] and BTEF [4]. With the focus on exploring the decision making process behind startup's behaviors, we attempted to apply the four principle of BTEF to explain for how do startups face with such technical challenges.

---

[1] www.goo.gl/r9okCu
[2] www.qsrinternational.com/product

## 4 How are theories of entrepreneurial behaviors applicable to explain for startup product development process?

Six identified themes were directly related to startup's decision making. We found that BTEF can be useful to explain for such themes, as shown in Table 1. The challenge name and description were given along with the theoretical explanation in the table.

**Table 1: Explanation for startup's technical challenges**

| **Challenges** | **Description** | **Interpretation via BTEF** |
|---|---|---|
| Vague prototype planning | Prototypes were created in an adhoc manner, mostly throw-away, lack of upfront design for learning, sometimes little lesson learnt, lack of early-stage product roadmap | Startups emphasizing short-run reaction rather than anticipation of long-run uncertain events focusing on the search for a suboptimal set of features or functionalities [4] |
| Feature creeps | Startups implement requirements from many customers with different needs, divergent product roadmap. "*We are adding features all the time. This is not a product that will ever stop evolving. … We are talking about this being the core of the company's competence*" | Startups tend to perform different experiments with technology, i.e. features, user experience etc. Many startup features are a good representation of technology of foolishness |
| Paradox of demonstration | Early demonstration needs to be impressive to attract funding. However there is often a limited budget for developing a minimum viable product | Startups operate based on mean-driven transformation [4]. Demonstrated prototypes were limited by the current human and financial resources |
| Evolving throw-away prototypes | Many throw-away prototypes accidentally become evolutionary ones. Technical debt caused by the lack of proper refactoring threatens the quality of product in later phases of software startups | Startups leverage contingencies [4]. Tolerating surprises during a series of prototypes might lead to utilize the business-fit prototype for long-term development |
| Sharing visions between Business and Technology | Communication of business or technical details can be difficult between entrepreneurs and developers. "*it always takes a long discussion to explain her [the CEO] about the importance* | Conflicts do not necessarily happen in a startup context, as startup team members are both persuadable and persuasive to different degrees about different matters [4] |
| | *of having flexible product design...*" | |
| Lack of sufficient and relevant user involvement | Balancing learning fast and learning the right things is a challenging task. Startups might have problems with finding feedbacks from relevant users. "*Most of them don't understand the idea … It probably came ten years before the app developers can recognize its benefit …*" | Challenges of early user involvement can be tracked to two problems, 1) to find appropriate early innovators and 2) whether there actually is a market for the product |

## 5 Discussions

The technical challenges were interpreted in a context of a prototype-centric development paradigm [13, 14, 15]. Literature reveals that startups adopt rapid releases to build a prototype in an evolutionary fashion and quickly learn from the users' feedback to address the uncertainty of the market. The rapid development approaches were found to improve the effectiveness of the requirement elicitation of any software development [15]. However, in many cases software startups do not throw away quick-and-dirty prototypes and evolve them (or part of them) into the final products.

By using an effectuation theory [4], we can explain different technical challenges that startups face with during their journeys from idea to commercialization. Technical challenges related to prototyping and product development are linked with startup's current capacity, experimental nature of technology development, risk tolerance and favor of short-run feedbacks. Driven by the existing means and resource, startups search for local optimal solutions, emphasize on short-run feedback rather than long-run strategies. This results in technical challenges, such as vague prototype planning, and paradox of demonstration. All in all, the theory suggests the observed product development approaches do not likely support the entrepreneurial processes or vice versa. Alternatively, the effectual decision making might need a better software development paradigm that can fit to the uncertain and dynamic situations of startups.

## 6 Conclusions

This paper portrayed six technical challenges in early phases of software startups. Entrepreneurs make decisions to search for local optimal solutions, emphasize on short-run feedback rather than long-run strategies, which might require a more proactive, flexible and agile approach than the state-of-practice software startup product development. Our contributions are two folds. Firstly, we illustrate for the application of firm's behavior theory in the relation to technical decisions, which are essential for achieving core values of software-based startups. Secondly, this is among the first attempt to bring a theoretical framework from entrepreneurship literature in Software Engineering. This is encouraging due to the current limited theoretical contribution to software startups research [13, 14].

There are several possibilities for future work on software startups. Our next step is to extend the map of startups challenge to include non-technical challenges that we identify from the cases, such as lock-in to external resources, changing team composition and market uncertainty. Furthermore, we found that entrepreneurial theories are helpful in understanding and explaining the context of technical challenges and decision-making. Future work would investigate more on how other theories can be adopted in software startup research.